**Article type: Communication**
**Title** Strong Ferromagnetism Achieved via Breathing Lattices in Atomically Thin Cobaltites

*Sisi Li,† Qinghua Zhang,† Shan Lin,† Xiahan Sang, Ryan F. Need, Manuel A. Roldan, Wenjun Cui, Zhiyi Hu, Qiao Jin, Shuang Chen, Jiali Zhao, Jia-Ou Wang, Jiesu Wang, Meng He, Chen Ge, Can Wang, Hui-Bin Lu, Zhenping Wu, Haizhong Guo, Xin Tong, Tao Zhu, Brian Kirby, Lin Gu, Kui-juan Jin,\* and Er-Jia Guo\**

S. S. Li, Dr. Q. H. Zhang, S. Lin, Q. Jin, S. Chen, Dr. J. Wang, M. He, Dr. C. Ge, Prof. C. Wang, Prof. H. B. Lu, Prof. T. Zhu, Prof. L. Gu, Prof. K. J. Jin, and Prof. E. J. Guo
Beijing National Laboratory for Condensed Matter Physics and Institute of Physics, Chinese Academy of Sciences, Beijing 100190, China
E-mail: kjjin@iphy.ac.cn and ejguo@iphy.ac.cn

S. S. Li and Prof. Z. P. Wu
State Key Laboratory of Information Photonics and Optical Communications and Laboratory of Optoelectronics Materials and Devices, School of Science, Beijing University of Posts and Telecommunications, Beijing 100876, China

Prof. X. Sang, W. Cui, and Dr. Z. Hu
State Key Laboratory of Advanced Technology for Materials Synthesis and Processing & Nanostructure research center, Wuhan University of Technology, 122 Luoshi Rd., Wuhan 430070, China

Dr. R. F. Need and Dr. B. Kirby
NIST Center for Neutron Research, National Institute of Standards and Technology (NIST), Gaithersburg, MD 20899, United States

Dr. R. F. Need
Department of Materials Science and Engineering, University of Florida, Gainesville, Florida 32611, United States

Dr. M. A. Roldan
Eyring Materials Center, Arizona State University, Tempe, AZ 85287, United States

S. Chen and Prof. H. Z. Guo
School of Physical Engineering, Zhengzhou University, Zhengzhou 450001, China

Dr. J. Zhao and Prof. J. O. Wang
Institute of High Energy Physics, Chinese Academy of Sciences, Beijing 100049, China

Prof. C. Wang, Prof. T. Zhu, Prof. L. Gu, Prof. K. J. Jin, and Prof. E. J. Guo
Songshan Lake Materials Laboratory, Dongguan, Guangdong 523808, China

Prof. Xin Tong and Prof. T. Zhu
China Spallation Neutron Source, Institute of High Energy Physics, Chinese Academy of Sciences, Beijing 10049, China

Prof. E. J. Guo
Center of Materials Science and Optoelectronics Engineering, University of Chinese Academy of Sciences, Beijing 100049, China

**Abstract**: Low-dimensional quantum materials that remain strongly ferromagnetic down to monolayer thickness are highly desired for spintronic applications. Although oxide materials are important candidates for next generation of spintronics, ferromagnetism decays severely when the thickness is scaled to the nanometer regime, leading to deterioration of device performance. Here we report a methodology for maintaining strong ferromagnetism in insulating $LaCoO_3$ (LCO) layers down to the thickness of a single unit cell. We find that the magnetic and electronic states of LCO are linked intimately to the structural parameters of adjacent "*breathing lattice*" $SrCuO_2$ (SCO). As the dimensionality of SCO is reduced, the lattice constant elongates over 10% along the growth direction, leading to a significant distortion of the $CoO_6$ octahedra, and promoting a higher spin state and long-range spin ordering. For atomically thin LCO layers, we observe surprisingly large magnetic moment (0.5 $\mu_B$/Co) and Curie temperature (75 K), values larger than previously reported for any monolayer oxide. Our results demonstrate a strategy for creating ultrathin ferromagnetic oxides by exploiting atomic heterointerface engineering, confinement-driven structural transformation, and spin-lattice entanglement in strongly correlated materials.

**Main text**

Functional materials with reduced dimensionality and excellent magnetic properties are desired for the next-generation spintronic devices that require smaller size and higher performance. Extensive effort has been invested to identify new materials that retain ferromagnetism with monolayer thickness. Recent work has demonstrated that the two-dimensional (2D) van der Waals (vdW) materials, such as $Fe_3GeTe_2$ and $CrI_3$, [1, 2] exhibit itinerant ferromagnetism with an out-of-plane magnetocrystalline anisotropy. However, key challenges preventing their widespread incorporation in spintronic devices remain, including 1) the relatively small magnetic moment, which can be further suppressed by thermal fluctuations according to the Mermin-Wagner theorem; [3] 2) the difficulty of growing large, high-quality, and stable-in-air single crystals for 2D devices; [4] and 3) the difficulty with integrating with CMOS technologies. Moreover, in many 2D vdW materials, electrons occupy the weakly correlated *s* and *p* orbitals that limit the ability to tune spin configurations by external fields.

Transition metal oxides (TMOs) with their large bandgaps (*insulating*), unpaired free electrons (*non-zero magnetic moment*), and strongly correlated interactions from *d* or *f* orbitals in the TM ions (*sensitive to various external fields*) provide an attractive pathway to generate thin-film ferromagnets. [5-8] Recently, the free-standing TMO films with a thickness down to monolayer limit attract great attention due to the relaxation of strain and chemically unbonded membranes. [9-11] Atomically thin functional oxides stimulate vast potential applications because they can be integrated with heterostructures of semiconductors and layered compounds. However, one key challenge remains that when TMO films are shrunk to dimensions below a critical thickness (typically four to five unit cells), the magnetic properties tend to deteriorate dramatically. [12-14] Typically, the ferromagnetic ordering temperature and saturation moment are greatly suppressed in ultrathin oxide layers, and an interfacial magnetic dead layer is formed, thus fundamentally limiting the use of TMOs in atomic-scale devices at present. [15-17]

Perovskite-type cobaltite, LaCoO$_3$ (LCO), is a ferromagnetic insulator in tensile strained thin films. [18-21] It has been studied intensively over last decade because the complete absence of power dissipation in a ferromagnetic insulator implies the lossless transfer of spin signals that circumvents the energy dissipation problem. [22, 23] In LCO, the delicate interplay between the crystal field splitting ($\Delta_{CF}$) and exchange interaction ($J_{ex}$) is extremely sensitive to strain. [20, 21] As a result, local structural modulation creates changes of Co-O bond length ($r_{Co-O}$) and Co-O-Co bonding angle ($\beta_{Co-O-Co}$), thus profoundly modifying the spin state of Co$^{3+}$ ions at the atomic scale, providing an opportunity to manipulate its magnetic ground state even shrinking to monolayer limit.

The copper oxide SrCuO$_2$ (SCO) has orthorhombic lattice symmetry with infinite-layer structure in bulk. [24] When the SCO is grown as a thin film, it forms a double-chain structure in a perovskite-like framework with missing apical oxygen. Upon reducing layer thickness, the planar-type (P-type) oxygen coordination structure of SCO transforms to a chain-type (C-type) structure to avoid a polar electrostatic instability that induces electronic reconstruction. [25] The CuO$_2$ planes are parallel (perpendicular) to the film plane in the P-type (C-type) SCO films. [26, 27] This structural transformation induces a large modification of out-of-plane lattice constant, oxygen coordination, and the crystallographic symmetry of SCO. Tuning this transition by choice of SCO layer thickness (Figure S1, Supporting Information) provides an opportunity to modify the local lattice distortion of adjacent layers in the heterostructures.

Here we synthesized superlattices composed of epitaxially grown LCO and SCO ultrathin layers. The thickness-dependent oxygen coordination transformation in SCO was exploited to directly perturb the structure of adjacent ferroelastic LCO at the atomic scale. We observed an enhanced magnetization in the LCO layers with a thickness down to single unit cell when the SCO layers exhibit a C-type structure. We corroborated these results with detailed analysis of octahedral parameters using scanning transmission electron microscopy (STEM) and unambiguously showed that the active spin state transition of cobalt ions triggered by octahedral distortion is responsible for the observed ferromagnetism in ultrathin LCO layers. Our work opens up opportunities for artificially designed spintronic devices based on atomically thin oxides.

The [LCO$_m$/SCO$_n$]$_{15}$ (L$_m$S$_n$) ($m$ = 1−5, $n$ = 1−20) superlattices were grown by pulsed laser deposition, where $m$ ($n$) denotes the number of unit cells (u. c.) of LCO (SCO) layers and 15 is the number of bilayer repetitions (See Experimental Section). Except for the first LCO layer, all LCO layers are sandwiched between two SCO layers of the same thickness in order to keep identical interfacial arrangement. X-ray diffraction (XRD) $\theta$-$2\theta$ scans and Reciprocal space mappings (RSMs) were conducted to comfirm the epitaxial, coherent growth of all superlattices (Figures S2 and S3, Supporting Information). Insight into the structural transformation occurring in our superlattices was obtained by high-resolution STEM studies. **Figure 1**a shows a cross-sectional high-angle annular dark-field (HAADF) STEM image from a L$_1$S$_8$ superlattice. The STEM image was acquired along the pseudocubic [100] zone axis. The intensity in the HAADF-STEM image scales roughly with the value of $Z^{1.7}$, where $Z$ is the atomic number of elements. The clear intensity contrast between lanthanum (La) and strontium (Sr) demonstrates that the ultrathin LCO layers with a thickness of one-unit-cell sandwiched between two SCO layers can be fabricated and controlled with unit cell precision. Figures 1b and 1c show the HAADF-STEM images of L$_3$S$_3$ and L$_3$S$_8$ superlattices, respectively, showing that the high degree of coherent structure can be maintained for a range of superlattices. The averaged atomic distances between the A-site cations (the brighter spots in HAADF-STEM images) along the [001] direction in the L$_3$S$_3$ and L$_3$S$_8$ superlattices are plotted in Figures 1d and 1e, respectively. We find the average out-of-plane lattice constant ($c$) of SCO is 3.80(5) Å for L$_3$S$_3$ and reduces to 3.45(3) Å for L$_3$S$_8$. Meanwhile, the $c$ of LCO increases from 3.78(4) Å for L$_3$S$_3$ to 3.80(2) Å for L$_3$S$_8$, which indicates a decrease in the tetragonality of LCO as SCO thickness increases. Similar results were observed in other L$_m$S$_n$ superlattice series (Figure S5, Supporting Information), providing additional evidence that the choice of SCO layer thickness generates a robust structural modification in the LCO layers. We notice that the out-of-plane lattice constants of SCO and LCO layers exhibit a difference over 10% between the adjacent layers. It is known that the lattice structure deforms from their bulk forms due to the substrate-induced epitaxial strain. The SCO layers are slight compressively strained (−0.53%), whereas the LCO layers are tensile-strained (1.7%). The out-of-plane strains of SCO and LCO layers are 0.52% and −1%, respectively. We calculate the Poisson ratio ($\varepsilon$) of SCO layers is 0.329 using its bulk lattice constants ($a = b = 3.926$ Å and $c = 3.432$ Å).[28] This value lies within the typical Poisson ratios reported in the most oxides with $\varepsilon$ ranging from 0.22-0.33,[29-32] suggesting the elastic deformation of SCO layers. Meanwhile, the $\varepsilon$ of LCO layers is determined to be 0.24, which is smaller than the values of LCO single films and bulk.[31, 32] We hypothesis that the formation of structural defects due to the growth of dissimilar materials may play a role in the relaxation of elastic energy, leading to a slight nonelastically structural deformation. Indeed, our STEM results confirm the occurrence of small number of structural defects in the films (Figure S4, Supporting Information). We perform the geometric phase analysis (GPA) on a typical STEM image. The strain states of superlattices vary slightly around these structural variations. In addtional, the chemical composition of $L_3S_3$ was ascertained using EELS mapping and profiling (Figure S6, Supporting Information). We find both interfaces between SCO and LCO layers do not exhibit significant intermixing with a maximum value ≈1 u.c. thick.

Annular bright-field (ABF) STEM images were collected in $L_3S_3$ and $L_3S_8$ superlattices to identify the oxygen coordination of SCO (Figure S7, Supporting Information). Local structural characterization shows directly the planar arrangement of oxygen atoms in the thick SCO layers ($n = 8$) changes into a chainlike structure in the thin SCO layers ($n = 3$). We also performed elemental-specific X-ray absorption spectra (XAS) measurements using linearly polarized X-rays to determine the electronic states of superlattices (Figure S8 and Supplementary Note 1, Supporting Information). The typical features at the Cu $L$-edges and Co $L$-edges confirm the valence states are $Cu^{2+}$ and $Co^{3+}$ in the superlattices, indicating that all layers are stoichiometric with negligible charge transfer at the interfaces and minor oxygen vacancies present in the films. Figures 1f and 1g show the x-ray linear dichroism (XLD) at the Cu *L*-edges from the $L_3S_3$ and $L_3S_8$ superlattices, respectively. In $L_3S_8$, the XLD shows large negative values, suggesting most holes occupy the $d_{x^2-y^2}$ orbital in the P-type SCO, which is consistent with orbital configurations observed in layer-structured cuprates.[33] In $L_3S_3$ for which the oxygen environment is more symmetric along in-plane and out-of-plane crystallographic directions, the XLD is close to zero, indicating holes are distributed equally over the $d_{x^2-y^2}$ and $d_{3z^2-r^2}$ orbitals. Together, our STEM and XLD results provide solid evidence of oxygen coordination transformation from P-type to C-type upon reducing the SCO layer thickness.

The intriguing oxygen coordination transformation in the SCO layers is accompanied by a change of magnetic ground states in the $L_mS_n$ superlattices. Although neither thick SCO nor ultrathin LCO films are ferromagnetic (Figure S9, Supporting Information), [24, 34, 35] ferromagnetic ordering is observed when the two components are integrated together in a superlattice. Magnetization (*M*) data as a function of magnetic field (*H*) for the $L_5S_n$ (*n* = 1−20) superlattices at 10 K are shown in **Figure 2**a. For *n* ≤ 5, the *M-H* curves exhibit well defined magnetic hysteresis loops, providing unambiguous evidence of ferromagnetism in the superlattices at low temperatures. The saturation magnetization ($M_{sat}$) reaches 106 emu/cm$^3$ for the $L_5S_1$ superlattice. Figure 2b shows the field-cooled *M* versus temperature (*T*) curves for the same superlattice series. All samples undergo paramagnetic-ferromagnetic transitions at Curie temperature ($T_C$) ≈ 85 K.[18-21] We plot $M_{sat}$ and the coercive field ($H_C$) of $L_5S_n$ superlattices as a function of *n* in Figures 2d and 2e. Both $M_{sat}$ and $H_C$ increase abruptly as *n* decreases (Region II) and show no significant changes for *n* ≤ 5 (Region I). We note that the *n* dependences of $M_{sat}$ and $H_C$ resemble that of $c_{SL}$. As shown in Figure 2c, $c_{SL}$ increases from approximately 3.4 Å (close to bulk value of SCO) [24] to 3.8 Å as *n* decreases from 20 to 5, demonstrating that the SCO layers undergo the expected abrupt structural transformation with unit cell elongating along the $c$-axis. When $n \leq 4$, $c_{SL}$ remains nearly constant. These observations are consistent with earlier work on (SrTiO$_3$/SrCuO$_2$) [26] and (LaNiO$_3$/SrCuO$_2$) [27] superlattices. We note that changes in $c_{SL}$ becomes gradual as the LCO thickness ($m$) increases from 1 to 5. This is attributed to the increasing volume fraction of LCO in the superlattices. Similar $n$ dependences of $M_{sat}$ and $H_C$ are observed in L$_3$S$_n$ and L$_1$S$_n$ ($n = 1-20$) superlattices as well, suggesting a common origin of enhanced ferromagnetism in L$_m$S$_n$ ($n \leq 5$) superlattices. These results thus point towards a strong correlation between ferromagnetism and structure of the superlattices.

To examine the layer-specific origin of magnetism in the L$_m$S$_n$ superlattices, we performed element-specific X-ray magnetic circular dichroism (XMCD) measurements as depicted in **Figure 3**a. The measured XAS and XMCD spectra for the Co and Cu $L$-edges at 10 K under a field of 5 T are shown in Figure 3b. The XMCD at the Co $L$-edges show a large negative response, whereas the XMCD of the Cu $L$-edge show no detectable dichroism. This provides direct evidence that the SCO layers contain zero magnetic moment and unmistakably identifies the origin of ferromagnetism in the superlattices as the LCO layers.

Polarized neutron reflectivity (PNR) (Figure 3c) and X-ray reflectivity (XRR) were used to quantify the structural and magnetic depth profiles of the L$_5$S$_1$ (with C-type SCO) and L$_5$S$_{10}$ (with P-type SCO) superlattices. The XRR data show distinct Bragg reflections corresponding to the superlattice ordering, allowing for precise determination of the electron density profiles and each superlattice's bilayer repeat thickness (Figure S10, Supporting Information). Figure 3d shows the fit spin-up ($R^+$) and spin-down ($R^-$) PNR data for sample L$_5$S$_1$ at 10 K under a field of 3 T, plotted with Fresnel normalization (top) and as spin asymmetry (bottom) against the wave vector transfer $q$ ($= 4\pi\sin\theta_i/\lambda$), where $\theta_i$ is the incident angle and $\lambda$ is the neutron wavelength. To fit models to this data, we assumed nuclear profiles consistent with Figure S8 and zero magnetization in the SCO layers as determined from XMCD. The resulting best-fit models are shown in Figure 3e and reveal $M_{LCO} \approx 130$ emu/cm$^3$ (0.86 $\mu_B$/Co) through the bulk of the superlattice, with significant reduction at both the substrate and surface interfaces associated with the correspondingly different boundary conditions. [36] The $L_5S_{10}$ superlattice is similar, but with a lower $M_{LCO} \approx 59$ emu/cm$^3$ (0.39 $\mu_B$/Co) (Figure S11, Supporting Information). These $M_{LCO}$ values are in excellent agreement with the previously described magnetometry measurements. Moreover, these results, which are not obfuscated by volume averaging over the non-magnetic SCO layers, conclusively show that $M_{LCO}$ is modulated by changes in SCO layer thickness and is different for the superlattices with P- and C-type SCO.

To understand the mechanism of the unexpectedly large magnetic moment observed in the ultrathin LCO layers, we performed the ABF-STEM imaging for $L_3S_3$ (**Figure 4**a) and $L_3S_8$ (Figure 4c) superlattices. The cross-sectional ABF-STEM images were acquired along the pseudocubic [110] zone axis to directly visualize the oxygen columns along with other heavy cations. We find that the oxygen ions sometimes not visible clearly in the $L_3S_3$ becasue the oxygen coordination of SCO is a typical C-type structure with oxygen plane either perpendicular or parallel to the projected direction. The viewzone may not perpendicular to the oxygen plane. The oxygen columns in the LCO layers are slightly below the Co plane by around 25(3) pm, suggesting an octahedral distortion with distinct suppressed $\beta_{Co-O-Co}$. This octahedral distortion in the LCO layers disappears in the $L_3S_8$ when the oxygen coordination of SCO changes to the P-type with oxygen plane aligning in the plane. Figures 4b and 4d show the layer-position-dependent bonding angle ($\beta_{M-O-M}$), where M represents the metallic ions (Ti, Co, Cu), in $L_3S_3$ and $L_3S_8$ superlattices, respectively. As illustrated by schematics of a single Co-O-Co bond, the $\beta_{Co-O-Co}$ averaged over the entire ABF-STEM images is 168(3)° for $L_3S_3$ and is close to 180° for $L_3S_8$. In $L_3S_3$, the octahedral modulations in the adjacent SCO is less than those of LCO and is absence in the STO substrate. Structural modulations in the LCO layers were also observed in the ABF-STEM images along the [100] zone axis (Figure S7, Supporting Information), further corroborating the existence of robust local structural distortions in LCO induced by the oxygen coordination transformation of the neighboring SCO. Please note that the displacements between the cations and anions suggest the possible existence of polar structure in LCO layers of $L_3S_3$ via interfacial structural engineering. [37, 38] We think the polar discontinuity and flexoelectric field induced by large strain gradient at the interfaces might be the reason for the lattice distortion in LCO layers.

The magnetization in LCO is directly linked to unconventional cobalt electronic states triggered by structural distortion. The local structural measurements show that LCO layers retain the octahedral oxygen coordination, as opposed to the tetrahedral oxygen coordination as the oxygen coordination in SCO (Supplementary Note 2, Supporting Information). In the pseudocubic octahedral ligand field, the Co-O molecular orbitals split into the threefold degenerate nonbonding $t_{2g}$ and twofold degenerate antibonding $e_g$ levels, as shown in Figure 4e. Earlier work on LCO demonstrated that the electronic spin state of $Co^{3+}$ ions is actively controlled by the energy difference $\Delta E = \Delta_{CF} - J_{ex} - W/2$, where $J_{ex}$ is an intrinsic materials constant, but $\Delta_{CF}$ ($\propto r_{Co-O}^{-5}$) and $W$ [$\propto r_{Co-O}^{-3.5} \cos(\pi - \beta_{Co-O-Co})$] strongly depend on the octahedral parameters, especially $r_{Co-O}$. In bulk LCO, the $\beta_{Co-O-Co} = 158º$ and $r_{Co-O} = 1.93$ Å,[39-41] and correspondingly the $Co^{3+}$ ions display a low-spin (LS) state at low temperatures. Since our LCO thin films are coherently strained to the underlying STO substrates, the atomic distance between Co ions is elongated by ≈ 1.1% relative to the LCO bulk value, and is a fixed value of √2$a_{STO}$/2 for both superlattices. The reduction of $\beta_{Co-O-Co}$ in $L_3S_3$ leads to ≈ 0.5% elongation of $r_{Co-O}$ in $L_3S_3$ compared to that in $L_3S_8$. Consequently, the $\Delta_{CF}$ reduces and the overall energy barrier for electrons jumping from $t_{2g}$ to $e_g$ levels decreases as well. The population of higher spin states (HS) $Co^{3+}$ ions increases and the Co-O bond covalency decreases, in agreement with direct XLD measurements on the electronic states of $Co^{3+}$ ions (Supplementary Note 1, Supporting Information). As a result, the spins favor ferromagnetic ordering, leading to a large magnetic moments in the ultrathin LCO layers.

The observed ferromagnetism in the our superlattices cannot be explained by chemical intermixing, i.e. Sr- or La-doping into LCO layers. First of all, the $L_1S_n$ superlattices show the clear $n$ dependence of $M$ (Figure 2d). The enhanced $M$ of $L_1S_n$, in consistent with other superlattice series, is observed when the SCO layers transit into C-type oxygen coordination. If the chemical doping plays a dominating role in the $L_mS_n$ superlattices, the $M$ should not depend on $n$. Secondly, the $M_{sat}$ of the $L_mS_1$ superlattices decays after $m > 5$ u. c. (Figure S12, Supporting Information). These results suggest there is a thickness limit for the enhanced $M$ in LCO layers by structural distortions. Specifically, we estimate the interfacial thickness of LCO layers with distorted octahedra is ~ 3 u. c. Increasing the LCO layer thickness in excess of twice the interface width does not appreciably enhance the $M$. [36] Thirdly, the $T_C$ of $L_mS_n$ is nearly constant ≈ 85 K, independent of $m$ or $n$. Previous work reveals that the $T_C$ of $La_{1-x}Sr_xCoO_3$ (0.01≤ $x$ ≤ 0.5) is above 200 K, [42-44] which is significantly higher than the $T_C$ of our superlattices. Therefore, we conclude chemical intermixing is unlikely to be a sufficient condition for the enhance $M$ in ultrathin LCO layers. Last but not the least, we demonstrate that the polar-nonpolar interfaces between LCO and SCO does not affect the electronic states of LCO layers, thus the systematic change in the superlattices' magnetization cannot be attributed to the polar discontinuity (Supplementary Note 3, Supporting Information).

Finally, we utilize this unique structural distortion imposed by oxygen coordination transformation to manipulate the magnetic state in the one-unit-cell-thick LCO layers. **Figures 5**a shows the HAADF-STEM image for a $L_1S_1$ superlattice. The alternate intensities from La and Sr ions together with the atomically resolved EELS intensities (Figure 5b) and intensity profiles (Figure 5c) indicate the ultra-sharp interfaces and small chemical intermixing across the alternative single layers. The ABF-STEM image shows the distinct octahedral distortion in LCO persists in the $L_1S_1$ superlattice (Figure S13, Supporting Information). $β_{Co-O-Co}$ is 173(2)° for $L_1S_1$, which is considerably smaller than that for $L_3S_8$. Large ferromagnetism, $M_{sat}$ ~75 emu/cm$^3$ (~ 0.5 $μ_B$/Co), was observed in the $L_1S_1$ superlattice (Figures 5d and 5e). The large magnetization follows in line with the spin state transition of Co ions. We think the small number of defects or dislocations in the sample may not dominate the overall magnetic ground state. As far as we know, this magnetic moment is the highest of any reported ferromagnetic TMO layers with a thickness of one-unit-cell. In addition, we observed a large magnetic anisotropy of the $L_1S_1$ superlattice with an easy-axis along the out-of-plane direction, similar to that of the monolayer 2D vdW materials, is favored because of the apparent crystallographic anisotropy. From the practical device applications, it is easier, faster, and more energy efficient to reverse an out-of-plane magnetization than the in-plane magnetization using the low-power optical pumping [45] and current-driven spin orbital torques (SOT). [46]

In summary, we report the observation of strong ferromagnetism in the ultrathin LCO layers with a thickness as thin as one-unit-cell. That the oxygen coordination of Cu ions in the SCO layers transforms upon reduction in thickness provides opportunities to modify the oxygen octahedral geometry of the adjacent LCO layers. When the SCO changes from P-type to C-type oxygen coordination, the oxygen octahedra in LCO distorts, promoting the high spin state cobalt ions, and greatly enhancing the magnetization in the ultrathin LCO layers. The methodology described in our work highlights the possibility of generating strong ferromagnetism in atomically thin TMO layers, which are highly desirable for the spintronic applications. Additionally, our local structural measurements indicate a distinct atomic displacement of cations with respect to the anions, implying a possible polar character in LCO ultrathin layer engineered by adjacent SCO. Strong coupling among different ferroic orders paves a new route towards atomically engineering the functionalities previously inaccessible or unexplored in these artificial-designed quantum heterostructures.

**Experimental Section**
***Synthesis of thin films and superlattices*** LaCoO$_3$(LCO) and SrCuO$_2$(SCO) single layers, and [LCO$_m$/SCO$_n$]$_{15}$ superlattices, where *m* and *n* are the numbers of unit cells, were grown on the TiO$_2$-terminated SrTiO$_3$ (STO) single crystalline substrates (Hefei Kejing Mater. Tech. Co., LTD) using pulsed laser deposition (PLD). The sintered LCO and SCO stochiometric targets were used for laser ablation. During the deposition, the oxygen partial pressure and substrate's temperature were maintained at 100 mTorr and 750 °C, respectively. The laser energy density was set at ~ 1.3 J/cm$^2$ and the laser repetition rate was 2 Hz. After the deposition, the samples were cooled down to room temperature at the oxygen partial pressure of 100 Torr. The growth rate for each component layer was calibrated carefully by x-ray reflectivity (XRR) using the thickness fringes from the whole layers' and satellites' reflections of the superlattices. XRR were fitted using GenX software. The accuracy of XRR data fitting is within one-unit-cell-thickness (± 4 Å). Examples of XRR fittings were shown in Figure S8. The calibrated growth rates are 23 pulses/u.c. for SCO and 45 pulses/u.c. for LCO.

*Structural and magnetization characterizations* XRD measurements for the single layers and superlattices were carried out on a Bruker D8 Discovery high-resolution diffractometer with Cu K$\alpha$1 radiation. Reciprocal space maps (RSMs) were taken around the substrates' 103 reflection. The sharp and clear Laue fringes and satellite peaks detected from superlattices in both RSMs and $\theta$-$2\theta$ scans attest to the smooth surface, abrupt interface, and highly epitaxial single crystalline layers. The macroscopic magnetization ($M$) of samples were measured by a Vibrating Sample Magnetometer (VSM, Quantum Design). The external magnetic field ($H$) was applied along the in-plane direction for all measurements. The $M$-$T$ curves were recorded during the sample warming-up after the field-cooling at $\mu_0 H$ = 1 kOe. The $M$-$H$ loops were taken at 10 K after the subtraction of diamagnetic signal from STO substrates. Since the SCO layers are nonmagnetic determined by VSM, PNR, and XMCD measurements, the magnetizations of all superlattices were normalized to the total thickness of LCO layers.

*Scanning transmission electron microscope (STEM)* Cross-sectional TEM specimens of superlattices were prepared using ion milling after mechanical thinning and precision polish. High-angle annular dark-field (HAADF) and annular bright-field (ABF) imaging and electron energy loss spectroscopy (EELS) analysis were carried out in a STEM. The L$_1$S$_n$ ($n$ = 3 and 8) superlattices were measured using a Nion UltraSTEM200 at Wuhan University of Technology (WUT). The L$_3$S$_n$ ($n$ = 3 and 8) and L$_1$S$_1$ superlattices were investigated using JEM ARM 200CF microscopy at Institute of Physics (IOP) of Chinese Academy of Sciences (CAS). The L$_5$S$_n$ ($n$ = 3 and 10) superlattices were studied using a JEM ARM 200F STEM at Eyring Materials Center (EMC) of Arizona State University (ASU). All STEMs operate at 200 kV and are equipped with double spherical aberration ($C$s) correctors for sub-angstrom resolution. Samples were imaged along the pseudocubic [100] zone axis for the analysis of chemical composition (HAADF) and the positions of the oxygen atoms (ABF). Images taken along the pseudocubic [110] zone axis were used for analyzing the octahedral distortion in the respective lattices. The atomic positions of La and Sr cations were determined by fitting the intensity peaks with Gaussian function over the large regions illustrated in the HAADF-STEM images. Error bars were extracted by calculating the standard deviation values from the measured distances between each atom. The elemental-specific EELS maps had been produced by integrating the signals from the interested regions after background subtracting using a power law. The STEM data were analyzed by Gatan Digital Micrograph software.

***Polarized neutron reflectometry (PNR)*** PNR experiment on $L_5S_1$ and $L_5S_{10}$ superlattices were performed at both MR beamline of the Chinese Spallation Neutron Source (CSNS) and the PBR beamline of the NIST Center for Neutron Research (NCNR). The samples were field-cooled and measured at $\mu_0 H = 3$ T. The magnetic field was applied along the in-plane direction, same as the VSM measurements. PNR measurements were carried out at 10 K in the specular reflection geometry with wave vector transfer ($q$) perpendicular to the surface plane. The neutron reflectivity was recorded as a function of $q$ for the spin-up ($R^+$) and spin-down ($R^-$) polarized neutrons. The neutron reflectivities in Figures 3 and S9 were normalized to the asymptotic value of the Fresnel reflectivity ($R_F = 16\pi^2/q^4$) for a better illustration. The difference between neutron reflectivities from $R^+$ and $R^-$ was illustrated by calculating the spin asymmetry [SA = $(R^+ - R^-)/(R^+ + R^-)$]. The PNR data were fitted using the chemical depth profiles obtained from XRR fittings, as shown in Figure S10. We used both GenX software and NIST Ref1D program to fit all PNR data. Both programs achieve the consistent fitting results.

***XAS and XMCD measurements*** Elemental-specific XAS measurements at the Co and Cu *L*-edges were collected at the beamline 4B9B of the Beijing Synchrotron Radiation Facility (BSRF). XAS were measured at room temperature in total electron yield (TEY) mode. X-ray linear dichroism (XLD) was characterized by changing incidence angle ($\alpha = 30°$ and $90°$) of linearly polarized x-ray beam. The incidence angle was varied by rotating the sample with respect to the direction of incoming x-rays. When the x-ray beam was perpendicular to the sample's surface ($\alpha = 90°$), it reflects the $d_{x^2-y^2}$ ($I_{ip} = I_{90°}$) orbital occupancy directly. When the x-ray beam incident on the sample's surface with an angle of $30°$, the XAS signal contains the information from both $d_{x^2-y^2}$ ($I_{ip} = I_{90°}$) and $d_{3z^2-r^2}$ [$I_{oop} = (I_{90°} - I_{30°}\cdot\sin^2 30°)/\cos^2 30°$] orbitals. The XLD can be simply calculated by $I_{ip} - I_{oop}$, as illustrated in Figures 1f and 1g. XMCD measurements were collected in TEY mode at 10 K with in-plane magnetic fields of $\mu_0 H = \pm 5$ T. XMCD were calculated from the difference between $\mu^+$ and $\mu^-$ divided by their sum, as described by $(\mu^+-\mu^-)/(\mu^++\mu^-)$, where $\mu^+$ and $\mu^-$ denote the XAS obtained from right-hand circular polarized and left-hand circular polarized x-rays, respectively.

***SHG measurements*** Optical SHG measurements were performed on the SCO single layers with different thicknesses. The SHG measurements were taken in the reflection geometry at room temperature. We used an 800 nm-wavelength laser from a Ti: Sapphire femtosecond laser (Tsunami 3941-X1BB, Spectra-Physics) as a pumping beam. The linearly polarized light incidents on the sample at an angle of 45°. The polarization direction ($\varphi$) of the incident field ($E_\omega$) was rotated through a $\lambda/2$ wave plate controlled by rotational motor. The second harmonic (SH) fields ($E_{2\omega}$) generated through the nonlinear optical process within the SCO films were decomposed into *p*- ($I_{2\omega\parallel}$) and *s*- ($I_{2\omega\perp}$) polarized components by a polarizing beam splitter, then were detected by a photon multiplier tube. Theoretical fittings of the SHG polarimetry data were performed with analytical models using standard point group symmetries.

**Supporting Information**

Supporting Information is available from the Wiley Online Library or from the author.

**Acknowledgements**

S. S. Li, Q. H. Zhang and S. Lin contributed equally to this work. We thank Z. Zhong, Z. Liao, T. Zhu, and M. R. Fitzsimmons for valuable discussions. This work was supported by the National Key Basic Research Program of China (Grants No. 2019YFA0308500), the National Natural Science Foundation of China (Grant Nos. 11974390, 51902237), the Beijing Nova Program of Science and Technology (Grant No. Z191100001119112), the Beijing Natural Science Foundation (Grant No. 2202060), and the Strategic Priority Research Program (B) of the Chinese Academy of Sciences (Grant No. XDB33030200). R.F.N. acknowledges support from the National Research Council Research Associateship Program. M.R. acknowledges the use of facilities within the Eyring Materials Center at Arizona State University. XAS experiments at Beijing Synchrotron Radiation Facility (BSRF) of the Institute of High Energy Physics, Chinese Academy of Sciences were conducted via a user proposal. The PNR measurements were conducted at the Chinese Spallation Neutron Source (CSNS) via a user proposal (P2018121100016) and at the National Institute of Standards and Technology (NIST) Center for Neutron Research (NCNR), U.S. Department of Commerce, via a quick access proposal. Any mention of specific trade names and commercial products is for information only; it does not imply recommendation or endorsement by NIST.

Received: ((will be filled in by the editorial staff))
Revised: ((will be filled in by the editorial staff))
Published online: ((will be filled in by the editorial staff))

**Figure 1. Structural and electronic state characterizations of the [(LaCoO$_3$)$_m$/(SrCuO$_2$)$_n$]$_{15}$ (L$_m$S$_n$) superlattices.** (a) Schematic (left) and HAADF-STEM image (right) of a L$_1$S$_8$ superlattice. (b) and (c) HAADF-STEM images of L$_3$S$_3$ and L$_3$S$_8$ superlattices, respectively. Samples are imaged along the pseudocubic [100] zone axis in cross-sectional view. (d) and (e) Atomic distances between the A-site cations along the [001] direction, i. e. out-of-plane lattice constants ($c$), in the L$_3$S$_3$ and L$_3$S$_8$ superlattices, respectively. The averaged $c$ values are obtained from the yellow rectangle regions marked in the HAADF images, and the errors bars represent one standard deviation. (f) and (g) X-ray linear dichroism (XLD) at the Cu $L$-edges for the L$_3$S$_3$ and L$_3$S$_8$ superlattices, respectively. Insets show the schematics of oxygen coordination transformation from a chain-type structure to a planar-type structure as the thickness of SrCuO$_2$ layers increases.

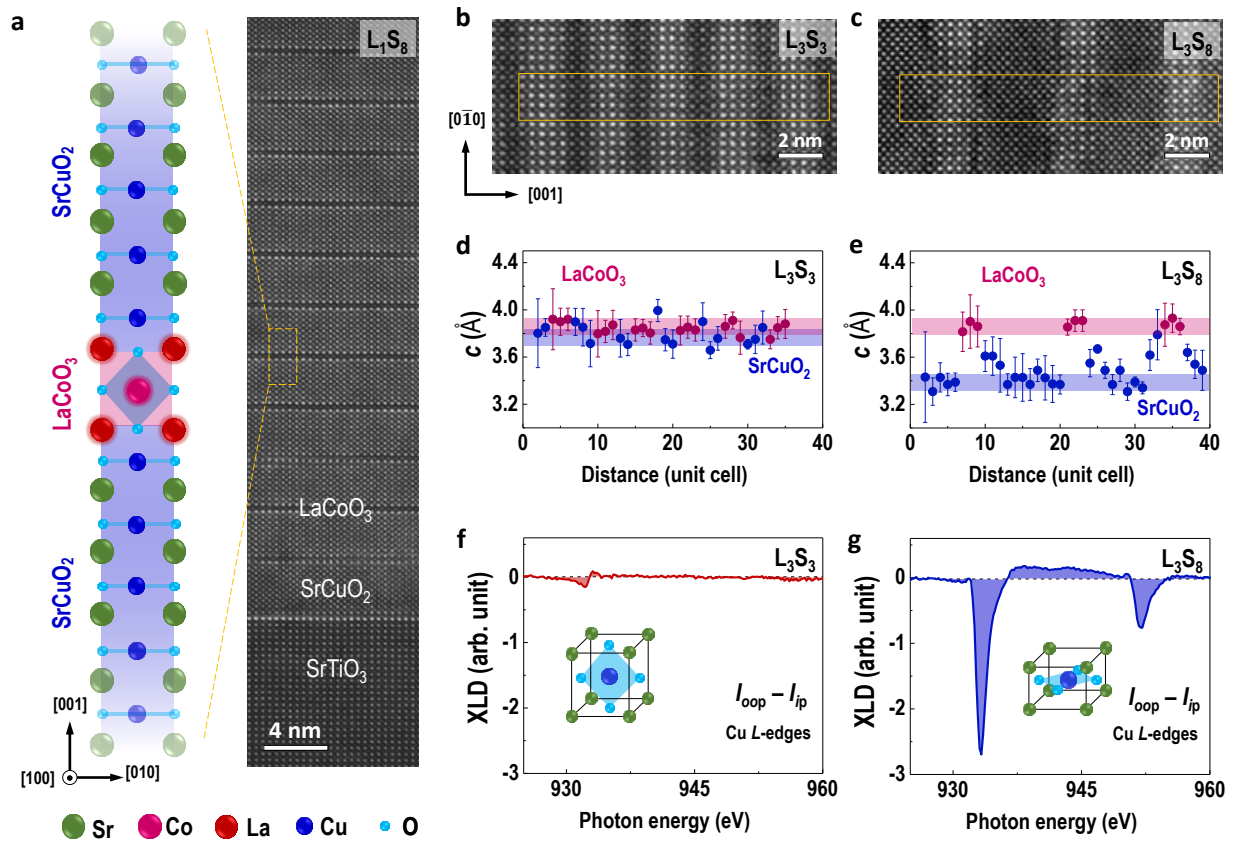

**Figure 2. Magnetic properties of superlattices controlled by the dimensionality of SrCuO₂ (SCO) layers.** (a) *M-H* loops and (b) *M-T* curves of $L_5S_n$ superlattice series for $1 \leq n \leq 20$, respectively. *M-H* loops were recorded at 10 K. *M-T* curves were measured after field-cooling at $\mu_0 H$ = 1 kOe. (c) Out-of-plane lattice constants ($c_{SL}$), (d) saturation magnetizations ($M_{sat}$), and (e) coercive fields ($H_C$) of $L_mS_n$ superlattices were plotted as a function of SCO layer thickness (*n*). Data from the superlattices with 1, 3, and 5 u. c.-thick LCO layers were summarized in (c)-(e). The errors bars represent one standard deviation. The horizontal dashed line in (c) indicates the out-of-plane lattice constant of bulk SCO. The $M_{sat}$ of superlattices were calculated using the total thickness of LCO layers (SCO layer is nonmagnetic, see Figure S7). Error bars are statistical error measurements. Region I and II indicate the dramatic change in the magnetic properties of superlattices corresponding to the oxygen coordination transformation in the SCO layers from a chain-type structure to a planar-type structure.

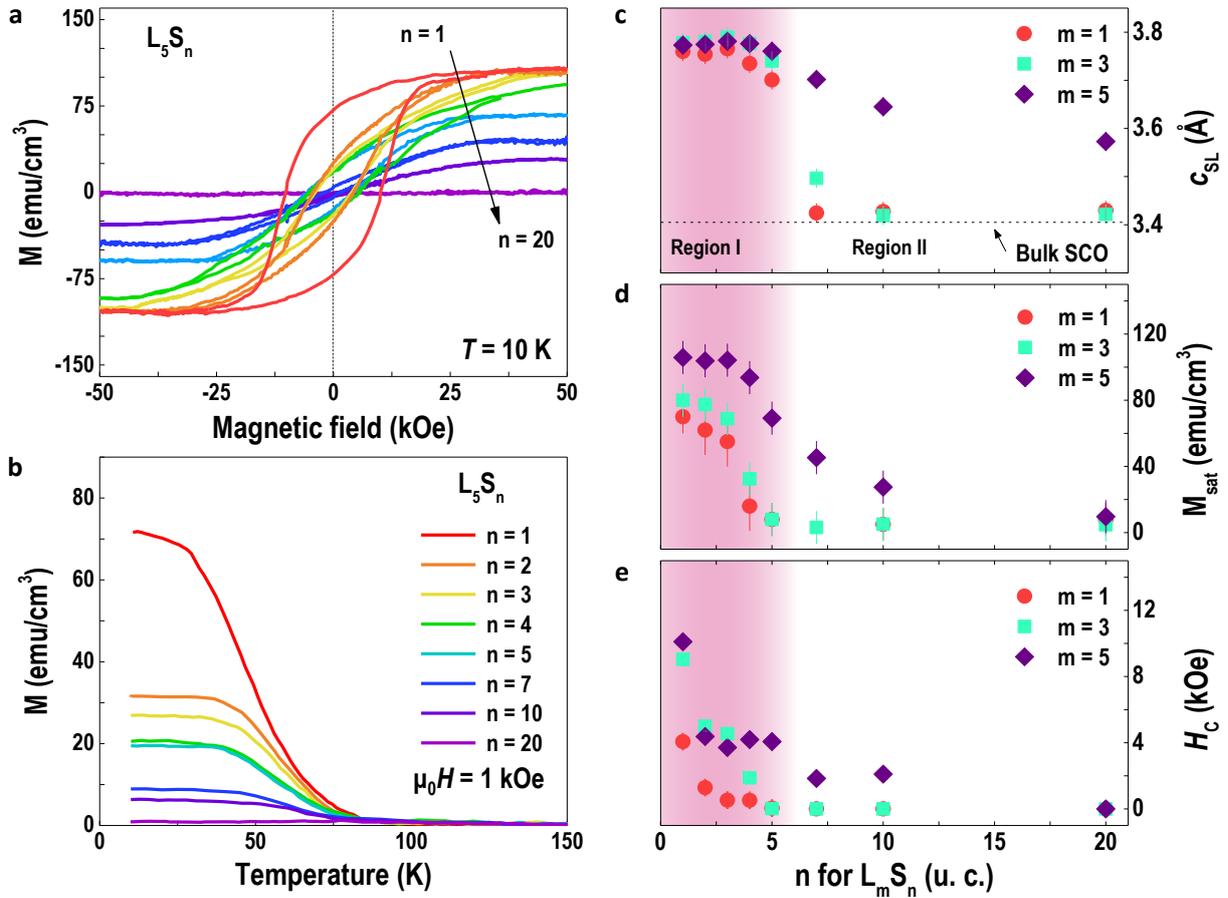

**Figure 3. Identifying the origin of magnetization in an [(LaCoO$_3$)$_5$/(SrCuO$_2$)$_1$]$_{15}$ superlattice by PNR and XMCD.** (a) and (c) schematic diagrams of experimental setups for XMCD and PNR measurements, respectively. (b) Elemental specific XAS and XMCD for Co and Cu $L$-edges. The reference spectra for Co$^{3+}$ [21] and Cu$^{2+}$ [27] were shown in black dashed lines for comparison. The presence of a shoulder at ~ 782 eV indicates that the LCO films substantially contain the low-spin state Co$^{3+}$ ions, in consistent with previous report [21]. The XMCD spectra were calculated from the difference between µ$^+$ and µ$^-$ divided by their sum, as described by (µ$^+$−µ$^-$)/(µ$^+$+µ$^-$), where µ$^+$ and µ$^-$ denote the XAS obtained from right-hand circular polarized and left-hand circular polarized x-rays, respectively. XAS was collected in the total electron yield (TEY) mode at 10 K with applied fields of $\mu_0 H = \pm 5$ T. (d) Reflectivity curves for spin-up ($R^+$) and spin-down ($R^-$) polarized neutrons are shown as a function of wave vector $q$ (upper panel). The reflectivity was normalized to the Fresnel reflectivity $R_F$ (=16π$^2$/q$^4$). PNR measurements were performed at 10 K after field-cooling at $\mu_0 H = 3$ T. The spin asymmetry (SA) was calculated by ($R^+$−$R^-$)/($R^+$+$R^-$) (lower panel). The errors bars represent one standard deviation. (e) Nuclear scattering length density (*n*SLD) and magnetic scattering length density (*m*SLD) depth profiles. The magnetization was calculated for a comparison to the magnetometry results. Inset of (e) shows the schematic drawing of the L$_5$S$_1$ superlattice geometry. Magnetic field was applied along the in-plane direction in both PNR and XMCD measurements.

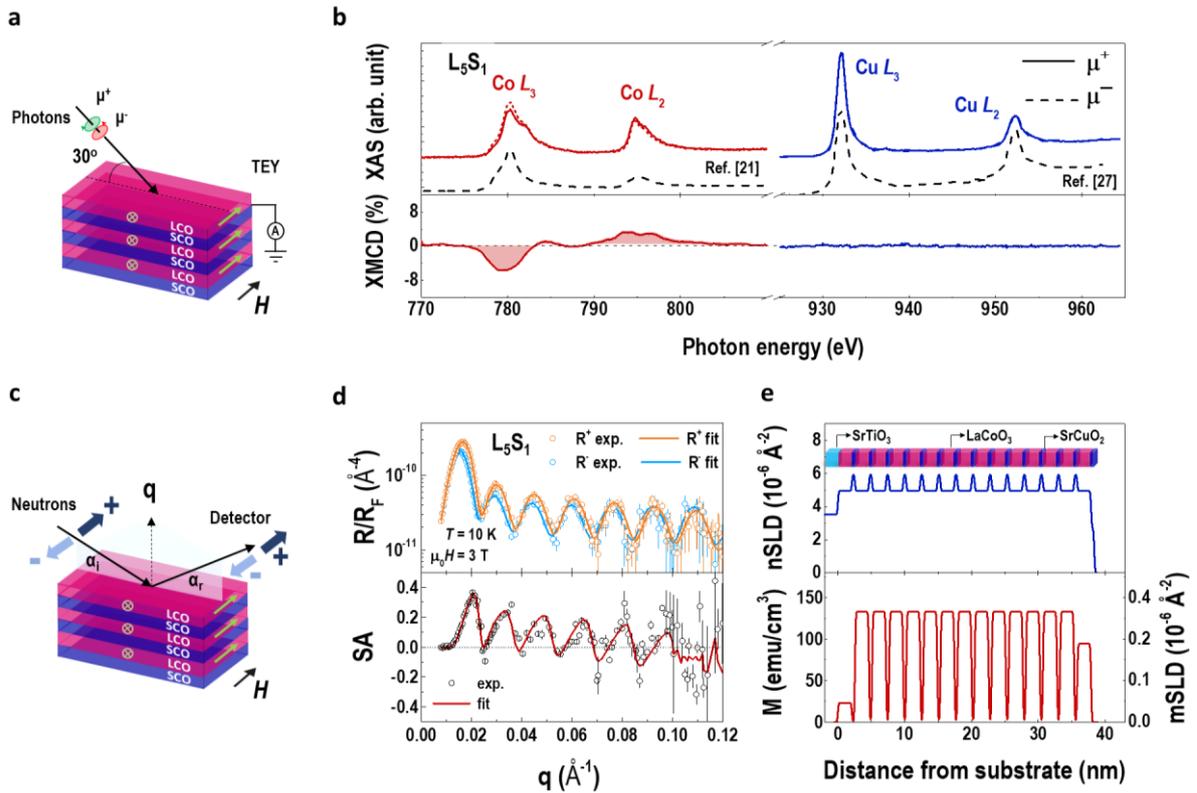

**Figure 4. Modifying the CoO$_6$ octahedral parameters by oxygen coordination.** (a) and (c) Cross-sectional ABF-STEM images of L$_3$S$_3$ and L$_3$S$_8$ superlattices, respectively. Samples were imaged along the pseudocubic [110] zone axis in cross-sectional view. (b) and (d) Layer-position-dependent bonding angle β$_{M-O-M}$ of L$_3$S$_3$ and L$_3$S$_8$ superlattices, respectively, where M represents metallic ions (Ti, Co, Cu). The β$_{M-O-M}$ is calculated by averaging the bonding angle from atomic planes within the images. Error bars represent one standard deviation. Statistical analysis indicates the mean bonding angle β$_{Co-O-Co}$ is ~ 168(3)$^o$ and ~ 179(1)$^o$ for the L$_3$S$_3$ and L$_3$S$_8$ superlattices, respectively. (e) Schematic energy-level diagrams of a Co$^{3+}$ ions with low-spin (LS), intermedia-spin (IS), and high-spin (HS) state configurations. The spin crossover between each spin state is controlled by the balance between the crystal-field energy (ΔCF) and e$_g$ band width, *i.e.* eventually corresponding to the change of the bond length ($r_{Co-O}$) and bonding angle (β$_{Co-O-Co}$) in the CoO$_6$ octahedra, respectively.

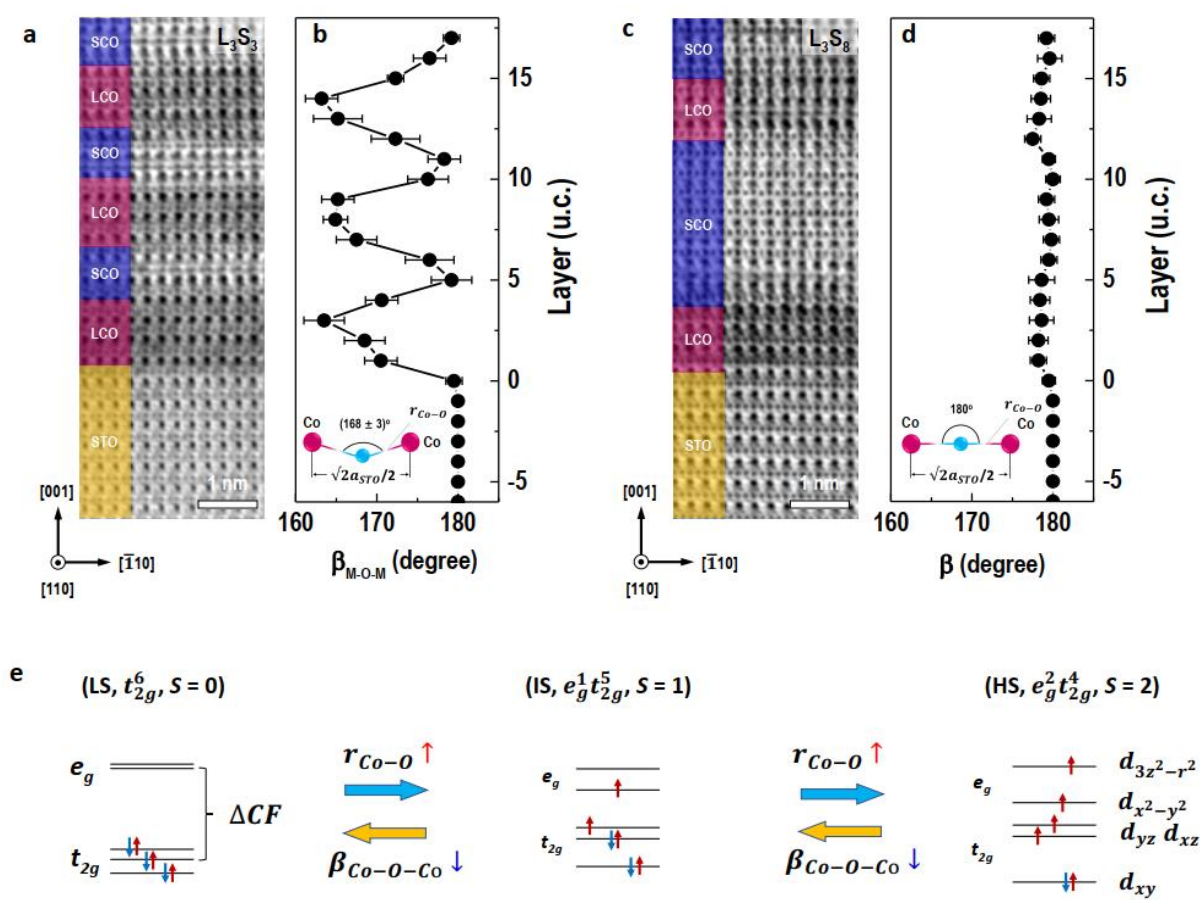

**Figure 5. Atomically engineered one-unit-cell LaCoO₃ layers yield a large magnetization.** (a) HAADF-STEM image of a selected area from the $L_1S_1$ superlattice, respectively. Samples were imaged along the pseudocubic [110] zone axis in the cross-sectional view. (b) Colored panels show the integrated EELS intensities of La-$M_{4,5}$, Co-$L_{2,3}$, Sr-$L_{2,3}$, Cu-$L_{2,3}$, and O $K$-edges, from which the unit-cell-precision of atomic layers can be resolved. (c) Intensity profile along the yellow line in (a), indicating the elemental positions in the atomic layers. (d) *M-H* loop and (e) *M-T* curves of a $L_1S_1$ superlattice, respectively. *M-H* loop was measured at 10 K. *M-T* curves were recorded during warming-up after field-cooling at 1 kOe. Both in-plane (red) and out of-plane (blue) magnetizations were measured.

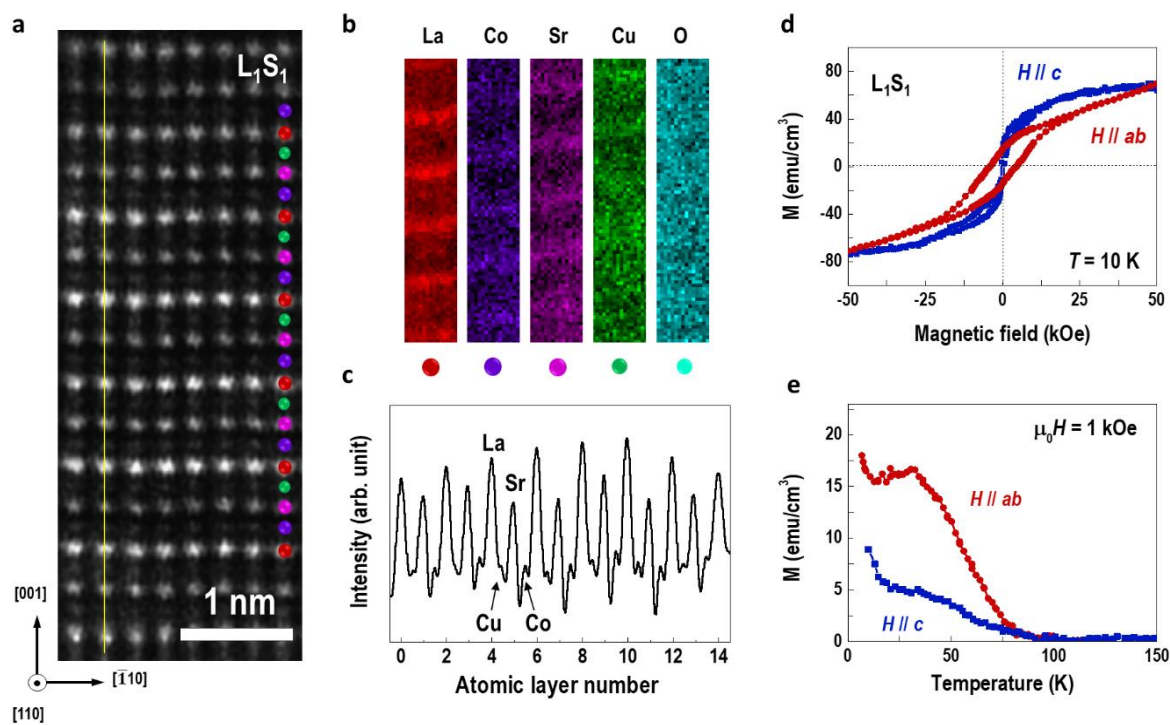